\documentclass[twocolumn,aps,prl,showpacs]{revtex4}

\usepackage[dvips]{graphicx}
\usepackage{amsmath}
\usepackage{SIunits}

\newcommand{\footnoteremember}[2]
{ \footnote{#2}
  \newcounter{#1}
  \setcounter{#1}{\value{footnote}}}

\begin{document}

\title{Entanglement of two individual neutral atoms using Rydberg blockade}
\author{T. Wilk, A. Ga\"{e}tan, C. Evellin, J. Wolters, Y. Miroshnychenko, P. Grangier, and A. Browaeys}
\affiliation{Laboratoire Charles Fabry, Institut d'Optique, CNRS, Univ Paris-Sud,
Campus Polytechnique, RD 128,
91127 Palaiseau cedex, France}

\date{\today}

\begin{abstract}
We report the generation of entanglement between two individual $^{87}$Rb atoms in hyperfine ground states $|F$=1,\,$M$=1$\rangle$ and $|F$=2,\,$M$=2$\rangle$ which are held in two optical tweezers separated by 4~$\mu$m. Our scheme relies on the Rydberg blockade effect which prevents the simultaneous excitation of the two atoms to a Rydberg state. The entangled state is generated in about 200~ns using pulsed two-photon excitation. We quantify the entanglement by applying global Raman rotations on both atoms. We measure that 61\% of the initial pairs of atoms are still present at the end of the entangling sequence. These pairs are in the target entangled state with a fidelity of 0.75.
\end{abstract}


\maketitle
Entanglement is at the heart of our understanding of quantum physics. It is useful as a resource for quantum information processing, as well as quantum metrology~\cite{Roos06} and can serve for the study of strongly correlated systems~\cite{Amico08}. In the last decades, entanglement has been demonstrated in many different systems such as photons~\cite{Aspect82}, ions~\cite{BlattWinelandNat08}, hybrid systems composed of an atom and a photon~\cite{Blinov04}, atomic ensembles~\cite{Julsgaard01,Chou05} and more recently between superconducting qubits~\cite{Steffen06}. Entanglement between two particles can be generated by designing and manipulating interactions between them, as it has been successfully demonstrated with trapped ions~\cite{BlattWinelandNat08}. It is more difficult to produce entanglement in neutral atom systems, due to their weaker interactions.
One solution, already implemented, makes use of a high-Q cavity to mediate the interaction between transient atoms~\cite{Hagley97}. Another recent approach uses ultra-cold atoms in optical lattices and s-wave interaction to create entangled states of many atoms~\cite{Mandel03, Anderlini07}. In this letter we demonstrate a different mechanism to entangle two neutral atoms in two hyperfine ground states, based on the Rydberg blockade. This approach has been proposed theoretically as a way to perform fast quantum gates~\cite{Jaksch00, Lukin01, Saffman05}. Recent proposals extended this idea to the generation of various entangled states~\cite{Moller08, Mueller09}. This approach is intrinsically deterministic and scalable to more than two atoms.

The principle of our scheme relies on the Rydberg blockade effect demonstrated recently with two single atoms~\cite{Urban09, Gaetan09}. Due to their large electric dipole when they are in a Rydberg state $|r\rangle$, atoms $a$ and $b$ interact strongly if they are close enough, and the doubly excited state $|r,r\rangle$ is shifted by $\Delta E$. As a consequence, a laser field coupling a ground state $|\!\uparrow\rangle$ and the Rydberg state $|r\rangle$ (with Rabi frequency $\Omega_{\uparrow r}$) cannot excite both atoms at the same time, provided that the linewidth of the excitation is smaller than $\Delta E$. In this blockade regime, the two-atom system absorbs only one excitation and behaves like an effective two-level system~\cite{Gaetan09}: the ground state $|\!\uparrow,\uparrow\rangle$ is coupled to the excited state
\begin{equation}\label{eqno1}
|\Psi_{\rm r}\rangle = \frac{1} {\sqrt{2}} (e^{i\mathbf{k}\cdot\mathbf{r}_a}|r,\uparrow\rangle + e^{i\mathbf{k}\cdot\mathbf{r}_{b}}|\!\uparrow,r\rangle),
\end{equation}
where $\mathbf{k}$ is related to the wave vectors of the excitation lasers and $\mathbf{r}_{a/b}$ are the positions of the atoms. The coupling strength between these states is enhanced by a factor $\sqrt{2}$ with respect to the one between $|\!\uparrow\rangle$ and $|r\rangle$ for a single atom~\cite{Gaetan09}. Thus, starting from $|\!\uparrow,\uparrow\rangle$, a pulse of duration $\pi/(\sqrt{2}\,\Omega_{\uparrow r})$ prepares the state $|\Psi_{\rm r}\rangle$. To produce entanglement between the atoms in two ground states, the Rydberg state $|r\rangle$ is mapped onto another ground state $|\!\downarrow\rangle$ using additional lasers (wave vector $\mathbf{k'}$, Rabi frequency $\Omega_{r \downarrow}$) with a pulse of duration $\pi/\Omega_{r \downarrow}$. This sequence results in the maximally entangled state
\begin{equation}\label{eqno2}
|\Psi\rangle = \frac{1} {\sqrt{2}} (|\!\downarrow,\uparrow\rangle + e^{i \phi}|\!\uparrow,\downarrow\rangle),
\end{equation}
with $\phi = (\mathbf{k} -\mathbf{k'})\cdot (\mathbf{r}_{b} - \mathbf{r}_{a}) $, assuming that the positions of the atoms are frozen (the moving atom case is discussed below). If the light fields are propagating in the same direction and the energy difference between the two ground states is small, $\mathbf{k}\simeq \mathbf{k'}$, we deterministically generate a well defined entangled state with $\phi =0$ which is the $|\Psi^+\rangle$ Bell state.

\begin{figure}
\begin{center}
\includegraphics[width=8.6cm]{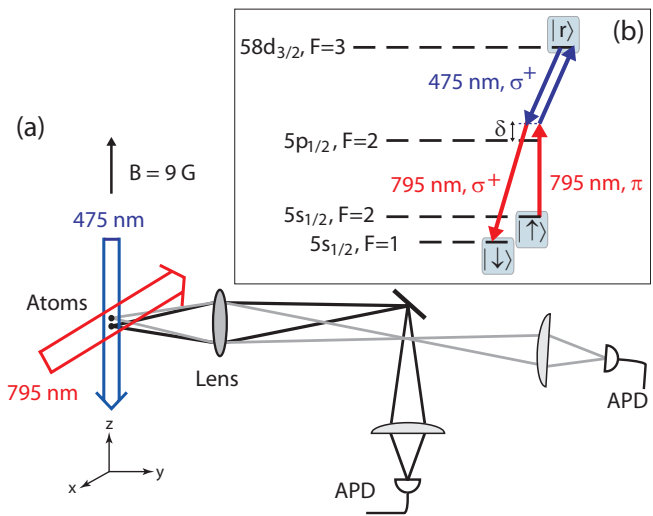}
\caption{Experimental setup (color online).
(a) Two atoms are held at a distance of 4~$\mu$m in two optical tweezers formed by focused laser beams at 810~nm (not shown). The fluorescence of each atom is directed onto separate avalanche photodiodes (APDs). The $\sigma^+$-polarized 475~nm laser has a waist of 25~$\mu$m and is directed along the z-axis, the two 795~nm lasers have waists of 130~$\mu$m, copropagate along the x-axis and have both linear polarization, one along the quantization axis, the other perpendicular. The 475~nm and 795~nm lasers have powers of 30~mW and 15~mW, respectively, which correspond to Rabi frequencies $\Omega_B/(2\pi) \sim 25$~MHz and $\Omega_R/(2\pi) \sim 300$~MHz.
(b) Atomic level structure and lasers used for the excitation towards the Rydberg state. The 475~nm laser and the two 795~nm lasers are tuned to the two photon transitions from $|\!\uparrow\rangle$ to $|r\rangle$ and from $|r\rangle$ to $|\!\downarrow\rangle$.}\label{figure1}
\end{center}
\end{figure}

Our experimental setup is depicted in Fig.~\ref{figure1}(a).
Two $^{87}$Rb atoms are held in two optical tweezers separated by 4~$\mu$m. The interatomic axis is aligned with a magnetic field ($B$=9~G), which defines the quantization axis and lifts the degeneracy of the Zeeman sublevels. The tweezers are formed by two laser beams at 810~nm which are sent at a small angle through a microscope objective focusing the beams to a waist of 0.9~$\mu$m. Atoms are captured from an optical molasses and, due to the small trapping volume, either one or no atom is captured in each trap~\cite{Schlosser01}. The same objective collects the fluorescence light of the atoms induced by the molasses beams at 780~nm. The light coming from each trapped atom is directed onto separate avalanche photodiodes (APDs) which allows us to discriminate for each trap whether an atom is present or not.

The relevant levels of $^{87}$Rb are shown in Fig.~\ref{figure1}(b). We have chosen the Rydberg state $|r\rangle$=$|58d_{3/2}$,\,$F$=3,\,$M$=3$\rangle$. The interaction energy between two atoms in this state is enhanced by a F\"{o}rster resonance~\cite{Walker08} which leads to a calculated interaction energy $\Delta E/h\approx 50$~MHz for a distance between the atoms of 4~$\mu$m~\cite{Gaetan09}. The ground states are $|\!\downarrow\rangle$= $|F$=1,\,$M$=1$\rangle$ and $|\!\uparrow\rangle$=$|F$=2,\,$M$=2$\rangle$ of the $5s_{1/2}$ manifold, separated in frequency by 6.8~GHz. To excite one atom from $|\!\uparrow\rangle$ to $|r\rangle$, we use a two-photon transition with a $\pi$-polarized laser at 795~nm and a $\sigma^+$-polarized laser at 475~nm. The frequency of the 795~nm laser is blue-detuned by $\delta$=600~MHz from the transition from $|\!\uparrow\rangle$ to ($5p_{1/2}$,\,$F$=2) in order to reduce spontaneous emission. The measured Rabi frequency of the two-photon transition from $|\!\uparrow\rangle$ to $|r\rangle$ is $\Omega_{\uparrow r}/2\pi\approx 6$~MHz for a single atom. We use the same 475~nm laser for the transition from $|r\rangle$ to $|\!\downarrow\rangle$, but a second 795~nm laser, linearly polarized perpendicular to the quantization axis, with a frequency 6.8~GHz higher to address state $|\!\downarrow\rangle$. The measured Rabi frequency for this second two-photon transition is $\Omega_{r\downarrow}/2\pi\approx 5$~MHz. The two 795~nm lasers are phase-locked to each other and can also be used to drive state rotations on the qubit states $|\!\uparrow\rangle$ and
~$|\!\downarrow\rangle$~\footnoteremember{foot1}{We observe Rabi
oscillations between $|\!\uparrow\rangle$ and $|\!\downarrow\rangle$ with an amplitude of 0.95, which includes the fidelity of state initialization and of state detection.}. The atomic state is read out by applying a push-out laser beam resonant to the $F$=2 to $F$=3 transition of the D2-line~\cite{Jones07}, which ejects atoms that are in state $|\!\uparrow\rangle$ (or in other $M$-states of the $F$=2 ground level) from the trap. Only atoms that are in $|\!\downarrow\rangle$ (or in other $M$-states of the $F$=1 level) will stay in the trap and will be detected.

An experimental sequence starts upon detection of an atom in each trap (trap depth 0.5~mK). After turning off the cooling beams, we optically pump the atoms in $|\!\uparrow\rangle$. Afterwards we switch off the dipole trap while we apply the excitation and mapping pulses towards the Rydberg state and back. The excitation pulse has a duration of $\pi/ (\sqrt{2}\,\Omega_{\uparrow r}) \approx 70$~ns to excite state $|\Psi_{\rm r}\rangle$. The mapping pulse has a duration $\pi/\Omega_{r\downarrow}\approx 110$~ns. Afterwards the trap is turned on again.
In order to analyze the produced two-atom state, we drive global Raman rotations on both atoms
(see below) before the push-out laser is applied.
Subsequently, we record for each trap whether the atom is present or not. From the results of 100 experiments we extract the probabilities $P_{a}$ and $P_{b}$ to recapture an atom in trap $a$ or $b$, the joint probabilities $P_{01}$ and $P_{10}$ to lose atom $a$ and recapture atom $b$ or vice versa, as well as probabilities $P_{11}$ and $P_{00}$ to recapture or lose both atoms, respectively.

We are interested in evaluating the amount of entanglement
between pairs of atoms present in the traps
at the end of the entangling sequence.
Our state detection scheme based on the push-out technique
identifies unambiguously any atom $a$ or $ b$ when it is in state $|\!\downarrow\rangle$.
However, it does not discriminate between atoms
in state $|\!\uparrow\rangle$ and atoms lost during the
sequence. In order to  count the
events for which both atoms are present at the end of the sequence, we
have measured in a separate experiment the probability to recapture a pair
of atoms after the excitation and mapping pulse, without applying the
push-out laser, $p_{\rm recap}=0.62(3)$
(the origin of these losses are detailed
below).  Only these remaining pairs, with correlations characterized by
the joint probability $P_{11}$, will be considered  for the analysis of
entanglement. This approach is very similar to the one used in Bell
inequality tests with {\it one-way polarizers}~\cite{Freedman72,Aspect81}.
In these experiments,
the number of relevant photon pairs is first measured by removing the
polarizers, and then only one qubit state is detected, while the other one
is lost.

We describe the final
two-atom state by the density matrix $\hat\rho$.
As a measure of entanglement we use the fidelity $F=\langle\Psi^+ |\hat\rho| \Psi^+\rangle = (P_{\downarrow\uparrow} + P_{\uparrow\downarrow})/2 + {\Re} (\hat\rho_{\downarrow\uparrow,\uparrow\downarrow})$ with respect to the expected $|\Psi^+\rangle$ Bell state~\cite{Sackett00} (${\Re }$ denotes the real part). The fidelity $F$ characterizes all the pairs present before the entangling sequence.  However, we are interested
in the fidelity $F_{\rm pairs}$ of the sample
consisting only of  the remaining pairs of atoms.
In the presence of one- or two-atom losses, we have ${\rm tr}(\hat\rho)<1$,
when restricting the trace to pairs of atoms still present at the end of the entangling sequence.
In this case the fidelity $F_{\rm pairs}$ is defined by
$F_{\rm pairs} = F/{\rm tr}(\hat\rho)$.

The  global Raman rotation allows
us to extract from $P_{11}$ all relevant diagonal and off-diagonal elements of $\hat\rho$ necessary to quantify the
amount of entanglement.
A Raman laser pulse of duration $t$, Rabi frequency $\Omega_{\uparrow\downarrow}$ and phase $\varphi$ rotates the state of each atom according to $|\!\uparrow\rangle \rightarrow \cos{\frac{\theta}{2}}|\!\uparrow\rangle + i e^{i\varphi}\sin{\frac{\theta}{2}} |\!\downarrow\rangle$, and $|\!\downarrow\rangle \rightarrow i e^{-i\varphi}\sin{\frac{\theta}{2}}|\!\uparrow\rangle + \cos{\frac{\theta}{2}} |\!\downarrow\rangle$ with $\theta=\Omega_{\uparrow\downarrow}\,t$.
The two atoms are exposed to the same laser field and undergo a rotation with the same angles $\theta$ and $\varphi$. In our experiment the phase $\varphi$ of the Raman pulse varies randomly from shot-to-shot over $2\pi$ and consequently our measurement results are averaged over $\varphi$. We calculate the expected values, averaged  over $\varphi$, of $P_{a/b}(\theta)$ and $P_{11}(\theta)$ as a function of the matrix elements of $\hat\rho$ and compare the expressions with the measurement results. When averaging over $\varphi$, all off-diagonal elements of $\hat\rho$ cancel in the observables we measure, apart from the relevant coherence $\hat\rho_{\downarrow\uparrow,\uparrow\downarrow}$.

The mean value of $P_{a/b}(\theta)$ is related to
the probability  $L_{a/b}$ to lose atoms $a$ and $b$, respectively,
after the excitation and mapping pulse. From
$\langle P_{a/b}(\theta)\rangle = \frac{1}{2} (1-L_{a/b})$ we find
$L_a=L_b = 0.22(1)$.
Assuming independent losses for atoms $a$ and $b$ we find the probability to lose at least
one of the two atoms $L_{\rm total}=L_a + L_b - L_a L_b = 0.39(2)$,
which is confirmed by
the recapture probability of a pair of atoms after the
excitation and mapping pulse
$p_{\rm recap}=0.62(3)$ quoted above.
Most of these losses are related to
the fact that an atom left in the Rydberg state is lost, since  it is not
trapped in the dipole trap. Using a model, we identify the following
mechanisms. Firstly, spontaneous emission from the $5p_{1/2 }$ state
populates the state  $|\!\downarrow\rangle$ from which $\sim 7~\%$ of the
atoms get excited to the Rydberg state by the mapping
pulse~\footnote{Spontaneous emission to state $|F=2,\,M=1\rangle$, which
would result in a loss ($\sim 2 \%$) from the qubit basis, is negligible
given our error bars,  due to a small branching ratio.}. Secondly,
intensity fluctuations ($5~\%$) and frequency fluctuations ($3$~MHz) of
the excitation lasers reduce the
efficiency of the mapping pulse so that $\sim 7~\%$ of the atoms will not
be transferred back from the Rydberg state to $|\!\downarrow\rangle$.
Independent of the Rydberg excitation, we measured
losses during the time the trap is switched off ($\sim 3\%$) as well as
errors in the detection of the presence of the atom ($\sim 3\%$).

\begin{figure}
\begin{center}
\includegraphics[width=8.6cm]{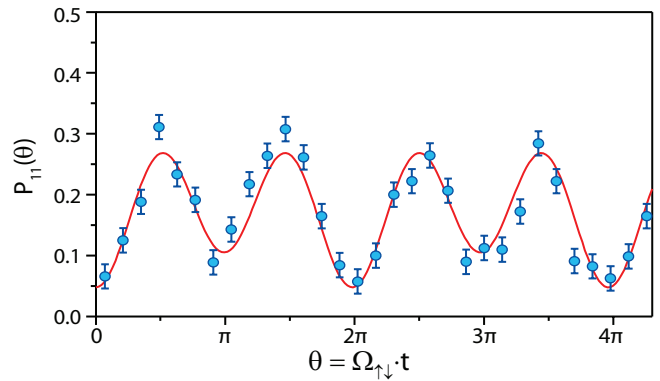}
\caption{Measured probability $P_{11}(\theta)$ to recapture the two atoms at the
end of the entanglement procedure, followed by a Raman pulse on both atoms for different
pulse durations. The data are fitted by a function of the
form $y_{0}+A\cos \Omega_{\uparrow\downarrow} t + B\cos 2 \Omega_{\uparrow\downarrow} t$,
according to the discussion in the text. The error bars on the data are statistical.}\label{figure2}
\end{center}
\end{figure}

The measurement result of $P_{11}(\theta)$ is shown in Fig.~\ref{figure2}. Ideally we expect $P_{11}(\theta)$ to oscillate between $0$ and $1/2$ with frequency $2\Omega_{\uparrow\downarrow}$ when varying the pulse duration $t$.
We observe indeed oscillations with frequency $2\Omega_{\uparrow\downarrow}$, but also oscillations with frequency $\Omega_{\uparrow\downarrow}$ which have an amplitude $(P_{\downarrow\downarrow} - P_{\uparrow\uparrow})/2$.
From the data we extract  $P_{\downarrow\downarrow} = P_{11}(0)$ and $P_{\uparrow\uparrow} = P_{11}(\pi)$.
Combining the value of the total losses $L_{\rm total}$
and the normalization condition  we get $P_{\uparrow\downarrow} +P_{\downarrow\uparrow}$.
The mean value  $\langle P_{11}(\theta)\rangle = (P_{\downarrow\uparrow} + P_{\uparrow\downarrow} +  3 P_{\downarrow\downarrow} + 3P_{\uparrow\uparrow} +2{\Re} (\hat\rho_{\downarrow\uparrow,\uparrow\downarrow})) / 8$ yields $\Re(\hat\rho_{\downarrow\uparrow,\uparrow\downarrow})$. Table~\ref{summary} summarizes the complete information about the density matrix $\hat\rho$ one can extract from global Raman rotations without control of $\varphi$.

\begin{table}
\begin{center}
\begin{tabular}{c c c}
\hline
\hline
Matrix elements & &Experimental values\\
\hline
$\hat\rho_{\downarrow\downarrow,\downarrow\downarrow}=P_{\downarrow\downarrow}$ & &$0.06\pm 0.02$\\
$\hat\rho_{\uparrow\uparrow,\uparrow\uparrow}=P_{\uparrow\uparrow}$ & &$0.09\pm 0.02$\\
$\hat\rho_{\downarrow\uparrow,\downarrow\uparrow}+\hat\rho_{\uparrow\downarrow,\uparrow\downarrow} =P_{\downarrow\uparrow}+P_{\uparrow\downarrow}$& &$0.46\pm 0.03$\\
${\Re}(\hat\rho_{\downarrow\uparrow,\uparrow\downarrow}) $& &$0.23\pm 0.04$\\
\hline
\hline
\end{tabular}
\caption{Measured values of the density matrix elements characterizing the state prepared in the experiment extracted from $P_{11}(\theta)$. The error bars are statistical. Note that ${\rm tr}(\hat\rho) = 0.61$ because of the loss $L_{\rm total} = 0.39(2)$ from the qubit states.}\label{summary}
\end{center}
\end{table}

Our analysis allows us to calculate the fidelity $F_{\rm pairs}$ of the sample
consisting only of the 61\% remaining pairs of atoms with respect to $|\Psi^+\rangle$.
Using $F=0.46$ and ${\rm tr}(\hat\rho) = 0.61 (\approx p_{\rm recap})$ as deduced from Table~\ref{summary},
we find $F_{\rm pairs} = 0.75(7)$ revealing the quantum nature of the observed correlations.
We emphasize that $F_{\rm pairs}$ is  the fidelity of the atom pairs,
which are present in the two traps before the state measurement.
It is therefore the amount of useful entanglement since it can be used for subsequent quantum
information protocols with pairs of atoms.
It is also the fidelity we would measure
if we had a detection scheme that could discriminate the internal states of an atom
which does not rely on expelling the atom from the trap.

We identify two mechanisms lowering the  fidelity $F_{\rm pairs}$ with respect to the ideal case. Firstly,
an imperfect Rydberg blockade which leads to the excitation of the two atoms
(probability $\sim 10\%$~\cite{Gaetan09}) and their subsequent mapping to
the state $|\!\downarrow,\downarrow\rangle$, leading to a non-zero
component of $P_{\downarrow\downarrow}$. Secondly,
the excess value of $P_{\uparrow\uparrow}$ is explained
by spontaneous emission and imperfect Rydberg excitation.

\begin{figure}
\begin{center}
\includegraphics[width=8.6cm]{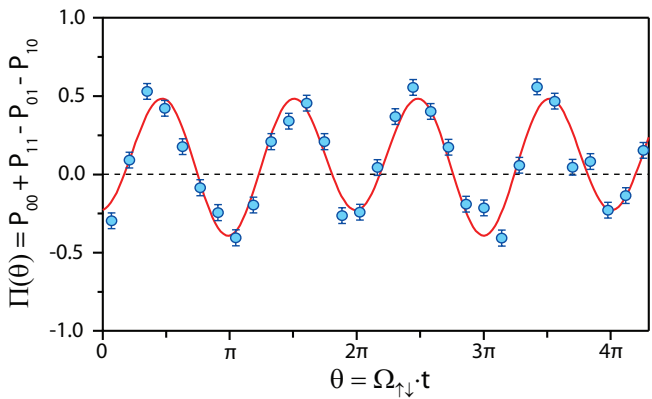}
\caption{Measured signal $\Pi$ for different durations of the analysing Raman pulse. The data are fitted by a function of the form $y_{0}+A\cos \Omega_{\uparrow\downarrow} t + B\cos 2 \Omega_{\uparrow\downarrow} t$ as discussed in the text. The error bars on the data are statistical.}\label{figure3}
\end{center}
\end{figure}

To confirm our results we analyze the signal $\Pi (\theta )= P_{00}+ P_{11}- P_{01}- P_{10}$ which is shown in Fig.~\ref{figure3} as a function of the Raman rotation angle $\theta$. The signal $\Pi (\theta )$ is equal to the parity~\cite{Turchette98} when there are no atom losses, then identifying a loss (0) with state $|\!\uparrow\rangle$ and a recapture (1) with state $|\!\downarrow\rangle$. For the maximally entangled state given by Eq.~\ref{eqno2}, the parity oscillates between $-1$ and $+1$ with a frequency of $2\Omega_{\uparrow \downarrow}$, while here the observed $\Pi (\theta )$ oscillates at two frequencies, $\Omega_{\uparrow\downarrow}$ and $2\Omega_{\uparrow\downarrow}$. The oscillation with $\Omega_{\uparrow\downarrow}$ can be assigned to events where one of the atoms is lost, while the oscillation with $2\Omega_{\uparrow\downarrow}$ can be attributed to cases where both atoms are present before the state detection. Then we calculate $\Pi (\pi/2) = 2{\Re} (\hat\rho_{\downarrow\uparrow,\uparrow\downarrow}) + L_a L_b$, from which we can deduce the coherence. We find ${\Re}(\hat\rho_{\downarrow\uparrow,\uparrow\downarrow}) = 0.22(4)$, which is in good agreement with the analysis described above, and shows once more the consistency of our data analysis.

So far we have neglected the influence of the atomic motion on the fidelity of the entangled state by assuming frozen motion of the atoms during the sequence. From an independent release and recapture experiment we measure a temperature of the atoms in the traps of $T=$60~$\mu$K~\cite{Tuchendler08}. Consequently, when the dipole trap is switched off during the entanglement sequence of duration $\delta t$, the atoms move in free flight. Hence, the phase in Eq.~\ref{eqno2} is now $\phi = \mathbf{k}\cdot(\mathbf{v}_{b}-\mathbf{v}_{a}) \delta t$, assuming $\mathbf{k}=\mathbf{k'}$, with $\mathbf{v}_{a/b}$ the velocity of the atoms. The phase $\phi$ fluctuates from shot-to-shot and the entanglement degrades with the strength of the fluctuations towards a mixed state. A simple model where we average the phase factor over the thermal velocity distribution yields $\langle e^{i\phi}\rangle = e^{-\Delta \phi^2/2}$, with $\Delta \phi = \sqrt{\frac{2 k_B T}{M} }|\mathbf{k}| \delta t$, where $M$ is the mass of the atom, and $k_B$ is the Boltzmann constant. For our shortest entangling sequence (pulse duration 70~ns and 110~ns, respectively, with 30~ns delay between the pulses), we calculate $\langle e^{i\phi}\rangle = 0.94$, resulting in a theoretical maximum fidelity $F_{\rm max} = 0.97$.

We repeated the experiment  with a separation of the excitation and mapping pulse of 600~ns. Following the procedure described above, we extract a decrease of ${\Re}(\hat\rho_{\downarrow\uparrow,\uparrow\downarrow})$ by a factor of 0.5 with respect to the case with 30~ns separation between the two pulses. This factor is in agreement with the theoretical value $\langle e^{i\phi}\rangle \approx 0.45$ given by our simple model. This indicates that the fidelity is ultimately limited by the residual temperature of the atoms, although this is presently not the main limiting factor.

In conclusion, we have demonstrated the entanglement of two atoms using the Rydberg blockade.
The 61\%  pairs of atoms  remaining at the end of the sequence are in a state with a  fidelity 0.75(7) with respect to
the expected $|\Psi^+\rangle$, showing the non-classical origin of
the correlations. Future work will be devoted to the measurement of the coherence time of the entangled state,
as well as the improvement of the fidelity and the state detection scheme.

We acknowledge the complementary work of the group from the University of Wisconsin \cite{Isenhower09}, where the authors demonstrate a CNOT gate based on the Rydberg blockade and use it to generate entangled states.

\begin{acknowledgments}
We thank P. Pillet, D. Comparat, M. M\"uller, M. Barbieri, R. Blatt, D. Kielpinski and P. Maunz for discussions and T. Puppe for assistance with the laser system. We acknowledge support from the EU through the IP SCALA, IARPA and IFRAF. A.\,G. and C.\,E. are supported by a DGA fellowship and Y.\,M. and T.\,W. by IFRAF.
\end{acknowledgments}


\begin{thebibliography}{40}
\bibitem{Roos06} C.\,F. Roos \textit{et al.}, Nature \textbf{443}, 316 (2006).

\bibitem{Amico08} L. Amico \textit{et al.}, Rev. Mod. Phys. \textbf{80}, 517 (2008).

\bibitem{Aspect82} A. Aspect, Nature \textbf{398}, 189 (1999).

\bibitem{BlattWinelandNat08} R. Blatt, and D. Wineland, Nature \textbf{453}, 1008 (2008).

\bibitem{Blinov04} B.\,B. Blinov \textit{et al.}, Nature \textbf{428}, 153 (2004).

\bibitem{Julsgaard01} B. Julsgaard, A. Kozhekin, and E.\,S. Polzik, Nature \textbf{413}, 400 (2001).

\bibitem{Chou05} C.\,W. Chou \textit{et al.}, Nature \textbf{438}, 828 (2005).

\bibitem{Steffen06} M. Steffen {\it et al.}, Science \textbf{313}, 1423 (2006).

\bibitem{Hagley97} E. Hagley {\it et al.}, Phys. Rev. Lett. \textbf{79}, 1 (1997).

\bibitem{Mandel03} O. Mandel {\it et al.}, Nature \textbf{425}, 937 (2003).

\bibitem{Anderlini07} M. Anderlini {\it et al.}, Nature \textbf{448}, 452 (2007).

\bibitem{Jaksch00} D. Jaksch {\it et al.}, Phys. Rev. Lett. \textbf{85}, 2208 (2000).

\bibitem{Lukin01} M.\,D. Lukin {\it et al.}, Phys. Rev. Lett. \textbf{87}, 037901 (2001).

\bibitem{Saffman05} M. Saffman, and T.\,G. Walker, Phys. Rev. A \textbf{72}, 022347 (2005).

\bibitem{Moller08} D. M{\o}ller {\it et al.}, Phys. Rev. Lett. \textbf{100}, 170504 (2008).

\bibitem{Mueller09} M. M\"{u}ller {\it et al.}, Phys. Rev. Lett. \textbf{102}, 170502 (2009).

\bibitem{Urban09} E. Urban {\it et al.}, Nature Phys. \textbf{5}, 110 (2009).

\bibitem{Gaetan09} A. Ga\"{e}tan {\it et al.}, Nature Phys. \textbf{5}, 115 (2009).

\bibitem{Schlosser01} N. Schlosser {\it et al.}, Nature \textbf{411}, 1024 (2001).

\bibitem{Walker08} T.\,G. Walker, and M. Saffman, Phys. Rev. A \textbf{77}, 032723 (2008).

\bibitem{Jones07} M.\,P.\,A. Jones {\it et al.}, Phys. Rev. A \textbf{75}, 040301 (2007).

\bibitem{Freedman72} S. J. Freedman, and J.\,F. Clauser, Phys. Rev. Lett. \textbf{28}, 938 (1972).

\bibitem{Aspect81} A. Aspect, P. Grangier, and G. Roger, Phys. Rev. Lett. \textbf{47}, 460 (1981).

\bibitem{Sackett00} C.\,A. Sackett {\it al.}, Nature \textbf{404}, 256 (2000).

\bibitem{Turchette98} Q.\,A. Turchette {\it et al.}, Phys. Rev. Lett. \textbf{81}, 3631 (1998).

\bibitem{Tuchendler08} C. Tuchendler {\it et al.}, Phys. Rev. A \textbf{78} 033425 (2008).

\bibitem{Isenhower09} L. Isenhower {\it et al.}, arXiv:0907.5552v3 (2009).

\end{thebibliography}
\end{document}